\documentclass[letterpaper,12pt]{article}   
\usepackage{ol2}  
\usepackage{amsmath}
\usepackage{bm}
\usepackage{amscd}
\usepackage{amssymb}
\usepackage{graphicx}
\providecommand{\abs}[1]{\left\lvert#1\right\rvert}

\begin{document}
%
%
\title{Brewster cross-polarization}
%
%
%
\author{A. Aiello,$^{1,2*}$  M. Merano,$^{1,}$  J. P. Woerdman$^{1}$}
\address{$^1$Huygens Laboratory, Leiden University,
P.O.\ Box 9504, 2300 RA Leiden, The Netherlands}
\address{$^2$ Max Planck Institute for the Science of Light, G\"{u}nter-Scharowsky-Str. 1/Bau 24, 91058 Erlangen, Germany}
\address{$^*$Corresponding author: aiello@molphys.leidenuniv.nl}
\begin{abstract}
We theoretically derive the polarization-resolved intensity distribution of a $TM$-polarized fundamental Gaussian beam reflected by an air-glass plane interface at Brewster incidence. The reflected beam has both a dominant ($TM$) and a cross-polarized ($TE$) component,
carried by  a $\text{TEM}_{10}$ and a $\text{TEM}_{01}$ Hermite-Gaussian spatial mode, respectively. Remarkably,  we find that the $TE$-mode power scales quadratically with the angular spread of the incident beam and it is comparable to the $TM$-mode power. Experimental confirmations of the theoretical results are also presented.
\end{abstract}




\maketitle 
When a beam of light impinges upon a plane interface separating
two transparent media, it produces reflected
and  transmitted beams.
In 1815 the Scottish physicist David Brewster discovered the total polarization
of the reflected beam at the angle $\theta_B$  since named after
him \cite{Brewster1815}. From his  observations he was also able to empirically determine the celebrated equation, known as Brewster's law, $\tan \theta_B = n_1/n_2$, where  $n_1$ and $n_2$ are the respective
refractive indices of the two media.
Several articles have been published on theory and experiments about Brewster's law for  beams with non-planar wave fronts.
Fainman and Shamir \cite{FainmanANDShamir} addressed reflection of an (isotropic) {spherical} wave, by a Brewster angle polarizer; they found a cross-polarized component. More recently, K\H{o}h\'{a}zi-Kis \cite{Kozaki} has theoretically derived and experimentally confirmed cross-polarization effects occurring at Brewster incidence.
Shortly afterwards, Li and Vernon \cite{LiANDVernon} addressed the same problem, using a microwave Gaussian beam; they do not mention cross polarization.
Consequences of cross-polarization coupling (XPC) at dielectric interfaces were   also theoretically investigated by Nasalski and coworkers in a series of interesting papers \cite{Nasalski01,NaPa}.

However, these studies fail to compare the intrinsic XPC due to the natural angular spread (namely, the focusing) of the incident beam, and the non-intrinsic one caused by reflection of such a beam at a dielectric interface.
 The main aim of this Letter is to fill this gap.

The structure of this Letter is as follows. We first  solve
 the general problem of the reflection of a polarized fundamental Gaussian beam at the plane interface between two optical media \cite{BliokhPRL,AielloOL2008}.  Next, we  derive analytical expressions for the polarization-dependent transverse spatial profiles of the reflected beam.
From this result, we are able to prove that non-intrinsic XPC scales quadratically with the angular spread $\theta_0$ of the incident beam, as opposed to the intrinsic XPC that scales as the fourth power of $\theta_0$.
Finally, we present experimental confirmations of our theoretical findings.

%
%
  \begin{figure}[!ht]
  \centerline{\includegraphics[width=14truecm]{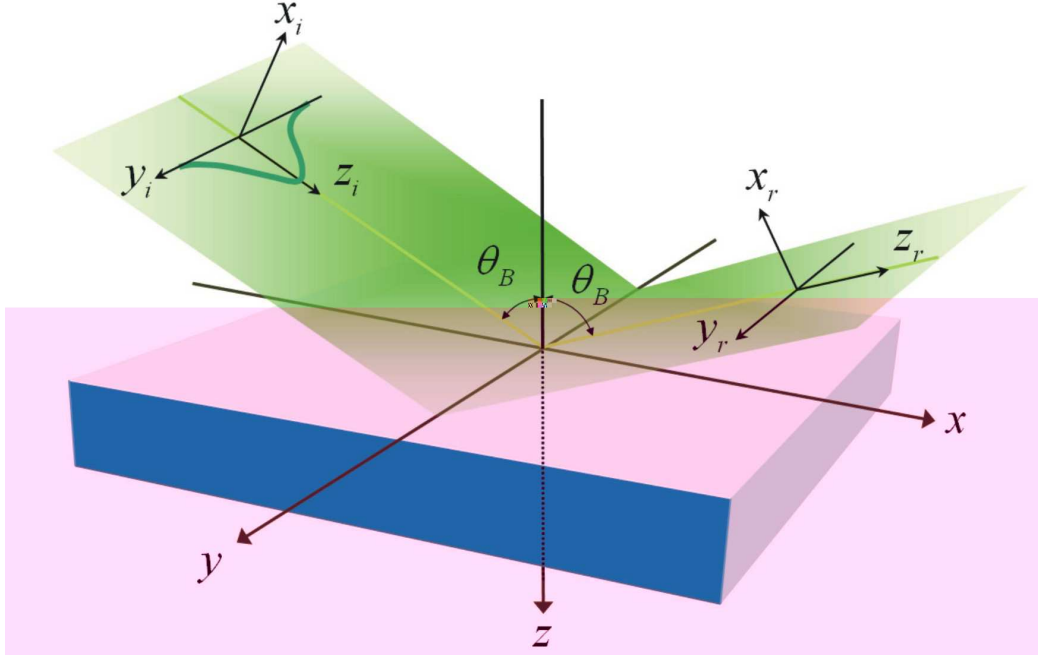}}
  \caption{(Color online) Geometry of beam reflection at the air-medium interface. $\theta_B$ is the Brewster angle.}
  \end{figure}
%
%
Consider a monochromatic beam of light incident upon a plane interface  that separates air from glass. With $n = n_\text{air}/n_\text{glass}$ we denote the ratio between the two refractive indices. 
As the beam meets the interface coming from the air side, it will be convenient to take the axis $z$ of the laboratory Cartesian frame $K=(O,x,y,z)$ normal to the  interface and directed from the air to the glass. Moreover, we choose the origin $O$ in a manner that the plane interface has  equation $z=0$.
The air-glass interface, the incident and the reflected beams are pictorially illustrated in Fig. 1. In addition to the laboratory frame,  we use a Cartesian frame $K_i=(O,x_i, y_i, z_i)$ attached to the incident beam and another one $K_r=(O,x_r, y_r, z_r)$ attached to the reflected beam.
Let $\mathbf{k}_0 = k_0 \hat{\mathbf{z}}_i$ and  $\mathbf{k}$ denote the central and noncentral wave vectors of the incident beam, respectively, with $\abs{\mathbf{k}} = \abs{\mathbf{k}_0} = k_0$.
Then, the electric field of the  incident beam  can be written  as a linear superposition of the fundamental vector plane-wave mode functions
 $\hat{\bm{\chi}}_\lambda(\mathbf{k})=\hat{\bm{e}}_\lambda(\mathbf{k}) \exp(i \mathbf{k}\cdot \mathbf{r} )$ with complex amplitudes  $a_\lambda(\mathbf{k})$, as follows:
\begin{align}\label{OL10}
\mathbf{E}^I(\mathbf{r} )
 = & \sum_{\lambda=1}^2 \int a_\lambda(\mathbf{k})  \hat{\bm{\chi}}_\lambda(\mathbf{k}) \,\text{d}^2 k_T,
\end{align}
 where ${\mathbf{k}}_T = \mathbf{k} - \mathbf{k}_0(\mathbf{k}_0 \cdot \mathbf{k})/k_0^2$  is the transverse part of $\mathbf{k}$, and we choose the polarization unit basis vectors   as $\hat{\bm{e}}_1(\mathbf{k}) = \hat{\bm{e}}_2(\mathbf{k}) \times {\mathbf{k}}/k_0$, and $\hat{\bm{e}}_2(\mathbf{k}) = \hat{\mathbf{z}} \times {\mathbf{k}}/\left|\hat{\mathbf{z}} \times {\mathbf{k}} \right|$
 \cite{NoteProducts}.
 Here $\hat{\mathbf{z}}$ is a real unit vector directed along the laboratory axis $z$.
By definition,  $\hat{\bm{e}}_1(\mathbf{k})$ lies in the plane of incidence  containing both the wave vector $\mathbf{k}$ and $\hat{\mathbf{z}}$, while $\hat{\bm{e}}_2(\mathbf{k})$ is orthogonal to such a plane.
A plane wave whose electric field vector is parallel to either  $\hat{\bm{e}}_1(\mathbf{k})$ or $\hat{\bm{e}}_2(\mathbf{k})$, is referred to as either  a \textsl{TM} or a \textsl{TE}  wave, respectively. The symbols  \textsl{S} for  \textsl{TE} and \textsl{P} for \textsl{TM}, are also widely used.
In Eq. (\ref{OL10})
$a_\lambda(\mathbf{k}) = A(\mathbf{k}) \, \alpha_\lambda(\mathbf{k})$, where
 $A(\mathbf{k})$ and $ \alpha_\lambda(\mathbf{k})$ are the scalar and the vector spectral amplitudes of the field, respectively.  
Here we consider a monochromatic Gaussian beam, whose spectral amplitude $A(\mathbf{k})$ is localized in $\mathbf{k}$ space, centered at the central wave vector $\mathbf{k}_0 = k_0 \hat{\mathbf{z}}_i$, on the sphere of equation $\omega^2(\mathbf{k}) = c^2 k_0^2$,  namely
\begin{align}\label{OL20}
A(\mathbf{k}) =   e^{ - \frac{\left|\mathbf{k}_T /k_0 \right|^2}{\theta_0^2}} e^{  i  k_0 d (1 -  \left|\mathbf{k}_T \right/k_0|^2 )^{1/2} } ,
\end{align}
where
$\theta_0 \equiv 2/(k_0 w_0)$ is the diffraction-defined angular aperture of the incident beam \cite{MandelBook} which has, by hypothesis, a minimum diameter (spot size) equal to $2 w_0$ located at $z_i = -d$.
The vector spectral amplitudes  are defined as   $\alpha_\lambda(\mathbf{k}) = \hat{\bm{e}}_\lambda(\mathbf{k}) \cdot \hat{\bm{f}}$, where $\hat{\bm{f}} = \bigl( f_P  \hat{\mathbf{x}}_i + f_S  \hat{\mathbf{y}}_i \bigr)$, with $|f_P|^2 + |f_S|^2=1$, is a complex-valued unit vector that fixes the polarization of the incident beam.

When the latter is reflected at the interface,  each vector  mode function changes according to
\begin{align}\label{OL30}
\hat{\bm{\chi}}_\lambda(\mathbf{k}) \mapsto r_\lambda(\mathbf{k}) \hat{\bm{\chi}}_\lambda({\widetilde{\mathbf{k}}}),
\end{align}
where $r_1(\mathbf{k})$ and $r_2(\mathbf{k})$ are the  Fresnel reflection amplitudes for \textsl{TM} and \textsl{TE} waves, respectively \cite{BandWBook}, and $ \widetilde{\mathbf{k}}= \mathbf{k} - 2 \, \hat{\mathbf{z}}\left(\hat{\mathbf{z}} \cdot \mathbf{k}  \right)$ is sets by the law of specular reflection  \cite{Gragg}.
If we substitute Eq. (\ref{OL30}) into  Eq. (\ref{OL10}), we obtain
\begin{align}\label{OL40}
\mathbf{E}^I (\mathbf{r} )  \mapsto  \mathbf{E}^R (\mathbf{r} )
 = & \sum_{\lambda=1}^2  \int   a_\lambda(\mathbf{k})r_\lambda(\mathbf{k}) \hat{\bm{\chi}}_\lambda({\widetilde{\mathbf{k}}}) \,\mathrm{d}^2 k_T,
\end{align}
where $\widetilde{\mathbf{k}}_0 = k_0 \hat{\mathbf{z}}_r$, by definition. The expression for the magnetic field  $\mathbf{B}^R(\mathbf{r} )$ of the reflected beam may be obtained from the equation above via the straightforward substitutions $a_\lambda(\mathbf{k}) \rightarrow b_\lambda(\mathbf{k})/c$, where $ b_1(\mathbf{k}) = - a_2(\mathbf{k}) r_2(\mathbf{k})/r_1(\mathbf{k})$, and $b_2(\mathbf{k}) =  a_1(\mathbf{k})r_1(\mathbf{k})/r_2(\mathbf{k})$.

From Eq. (\ref{OL20}) it follows that $A(\mathbf{k}) \simeq 0$ for those wave vectors $\mathbf{k}$ lying outside the paraxial domain $\mathcal{P} = \{\mathbf{k} : |\mathbf{k}_T/k_0| \lesssim \theta_0   \}$, with $\theta_0 \ll 1$ for well-collimated beams. This allows us to
 calculate analytically $\mathbf{E}^R$ (and, similarly, $\mathbf{B}^R $)  via a power series expansion for the integrand of Eq. (\ref{OL40}) about the point $\mathbf{k} = \mathbf{k}_0$  up
to and including second order terms in $\mathbf{k}_T/k_0$. In practice, we extend to the problem at hand, the perturbative approach introduced by Lax \emph{et al.}, \cite{LaxEtAl}, and further developed by Deutsch and Garrison \cite{DandG}.
It is easy to see that if with  $ \gamma(\mathbf{k}) = \arccos (\mathbf{k} \cdot \mathbf{k}_0/k_0^2)$ we denote the  angle between the central wave vector $\mathbf{k}_0$ and the non-central one $\mathbf{k}$, then we can write $|\mathbf{k}_T  /k_0| = \sin \gamma(\mathbf{k}) \simeq \gamma(\mathbf{k})$, where $ \gamma(\mathbf{k}) \lesssim \theta_0 \ll 1$,  being $\theta_0$ the natural small parameter for the power series expansion \cite{DandG}.
Explicit expressions for the power series expansions of both $\mathbf{E}^R(\mathbf{r} )$ and $\mathbf{B}^R(\mathbf{r} )$ are given in Appendix A.

From the knowledge of both $\mathbf{E}^R(\mathbf{r} )$ and $\mathbf{B}^R(\mathbf{r} )$ it is possible to calculate the intensity  spatial distribution (i.e., the beam profile) $I(\mathbf{r})$ of the reflected beam, as the flux of the cycle-averaged Poynting vector $\bar{\mathbf{S}} \propto \text{Re} \bigl(  \mathbf{E}^R \times {\mathbf{B}^R}^* \bigr)$ across a surface perpendicular to the central direction of propagation $\hat{\mathbf{z}}_r$, namely $I(\mathbf{r}) \propto \bar{\mathbf{S}} \cdot \hat{\mathbf{z}}_r$. Since $  \bigl(  \mathbf{E}^R \times {\mathbf{B}^R}^* \bigr) \cdot \hat{\mathbf{z}}_r =   E^R_{x_r}  {B^R_{y_r}}^* - E^R_{y_r}  {B^R_{x_r}}^* $, we can write $I(\mathbf{r}) = I_P(\mathbf{r}) + I_S(\mathbf{r})$, where $I_P(\mathbf{r}) = \text{Re} \bigl(  E^R_{x_r}  {B^R_{y_r}}^*\bigr)$, and $I_S(\mathbf{r}) = \text{Re} \bigl(-  E^R_{y_r}  {B^R_{x_r}}^*\bigr)$ are the intensity distributions produced by the $P$- and $S$-polarized components of the reflected beam, respectively. Here $P$ and $S$ polarization directions are defined with respect to the central plane of incidence containing  $\hat{\mathbf{z}}$,  ${\mathbf{k}}_0$, and $\widetilde{\mathbf{k}}_0$.
After a lengthy but straightforward calculation it is not difficult to obtain, for a $P$-polarized incident beam (i.e., for the choice $f_P =1$ and $f_S=0$), the following  power series expansions:
\begin{align}\label{OL50}
{I_P(\mathbf{r})}/{I_0(\mathbf{r})} \!  & = \!  r_P^2 + \theta_0 u X + \theta_0^2 \left( v + p X^2 + q Y^2\right), \\ \label{OL60}
{I_S(\mathbf{r})}/{I_0(\mathbf{r})} \!  & =     \theta_0^2 \, s Y^2,
\end{align}
where $X = x_r/w_0, \, Y = y_r/w_0, \, Z = (z_r + d)/L,$ and $I_0(\mathbf{r}) = \exp\left(-2\frac{X^2 + Y^2}{ 1 + Z^2} \right)/(1 + Z^2)$ is the  intensity distribution of the incident beam. In Eq. (\ref{OL50}) the transverse coordinates $x_r$ and $y_r$ are normalized with respect to the beam waist $w_0$, while the longitudinal coordinate $z_r$ is  normalized with respect to the Raleigh range $L = k w_0^2/2$ of the beam. In Eqs. (\ref{OL50}) and (\ref{OL60}), $r_P$ and $r_S$ are the Fresnel reflection amplitude for $P$ and $S$ waves, respectively, evaluated at the central angle of incidence $\theta = \arccos(\mathbf{k}_0 \cdot \hat{\mathbf{z}}/k_0)$, while $u, \, v, \,  p , \, q$ and $ s $ are some complicated functions of $Z, \theta, \,r_P, r_S$ and their derivatives \cite{NoteGH} whose explicit form is given in Appendix B.

However, at the Brewster angle of incidence $\theta = \theta_B \equiv \arctan(n)$, only  $ p $ and $ s $  take a non-zero value, namely
\begin{align}\label{OL70}
 p(1 + Z^2)  = \left.\bigl( \partial r_P/\partial \theta  \bigr)\right|_{ \theta_B}^2,  \;  s(1 + Z^2)  =  \left.{r_S^2}/{n^2}\right|_{\theta_B},
\end{align}
and Eqs. (\ref{OL50}) and (\ref{OL60}) reduces to
\begin{equation}\label{OL80}
{I_P(\mathbf{r})}/{I_0(\mathbf{r})}= \theta_0^2 \, p X^2, \qquad  {I_S(\mathbf{r})}/{I_0(\mathbf{r})} = \theta_0^2 \, s Y^2,
\end{equation}
  respectively. From this result it immediately follows that the ratio $\rho$ between the power of the $S$ and the $P$  components of the reflected beam at Brewster incidence, is simply equal to $ s / p $,
\begin{equation}\label{OL90}
\rho = \frac{{\iint I_S(\mathbf{r}) \text{d} X \text{d} Y }}{{\iint I_P(\mathbf{r}) \text{d} X \text{d} Y }} =   \left( \frac{r_S}{n} \frac{1}{\partial r_P/ \partial \theta}\right)_{\theta = \theta_B}^2.
\end{equation}
This simple result is remarkable since it shows that  $\rho$ is \emph{independent} from the waist $w_0$ of the incident beam, that is, $\rho$ take the same value for either a well-collimated or a strongly-focussed beam.

Equations (\ref{OL50}-\ref{OL60}) represent the main theoretical result of this Letter.
 In particular, Eq. (\ref{OL60}) shows that non-intrinsic XPC generates an intensity $I_S(\mathbf{r})$ that scales quadratically with $\theta_0$. This behavior cannot be ascribed to the intrinsic XPC exhibited by the incident beam since the latter is due to the beam-divergence  only and the consequent cross-polarized intensity scales with $\theta_0^4$ \cite{SimonANDMukunda87,EandS}.
Moreover, Eq. (\ref{OL80}) shows two important things. First, we see that even at Brewster angle of incidence the extinction of  a $P$-polarized beam is not perfect, as $\left.(d r_P/d \theta)\right|_{\theta_B} \neq 0$ and $\left. r_S\right|_{\theta_B} \neq 0$.
Thus, although the input beam is $P$-polarized, a $S$ component appears after reflection.
Second, the two expression in Eq. (\ref{OL80}) show that after reflection the cylindrical symmetry about the axis of propagation of the beam is lost and two orthogonally polarized $\text{TEM}_{10}$ and  $\text{TEM}_{01}$ modes are generated. The fact that  $I_S$ has a $\text{TEM}_{01}$ profile, as opposed to the cloverleaf $\text{TEM}_{11}$ pattern typical of the cross polarization intensity of the incident beam, is consistent with the hypothesis that such a term originates from non-intrinsic XPC  and it is not a simple beam-divergence effect.

We verified these theoretical results in our laboratory by using a Super-luminescent Light Emitting Diode (SLED) operating at $\lambda=820 \, \text{nm}$ (InPhenix IPSDD0802)  as a light source.
 The output of the SLED was first spatially filtered  by a single-mode optical fiber to prepare the input beam into the fundamental Gaussian mode, and then collimated  by a microscope objective to produce a very large beam waist, ($w_0 = 1.64 \, \text{mm}$) before passing across a polarizer selecting $P$-polarization.  A second lens put behind the polarizer generated the desired  waist for the input beam.
 The so-prepared beam was sent upon the surface of a right-angle BK7 glass ($n=1.51$)  prism mounted on a precision rotation stage with a resolution of $9 \times 10^{-6} \text{rad}$  (Newport URS-BCC) to accurately determine the Brewster angle ($\theta_B \cong 56.49^\circ$).
 Finally, the polarization-dependent beam intensity profiles after reflection were recorded by a CCD-based Beam Intensity Profiler (Spiricon LBA-FW-SCOR-$20$)  mounted behind a polarizer put along the axis of the reflected beam, at large distance from the interface (far field measurement).
%
%
%
  \begin{flushright}
  \begin{figure}[!hb]
  \centerline{\includegraphics[width=10truecm]{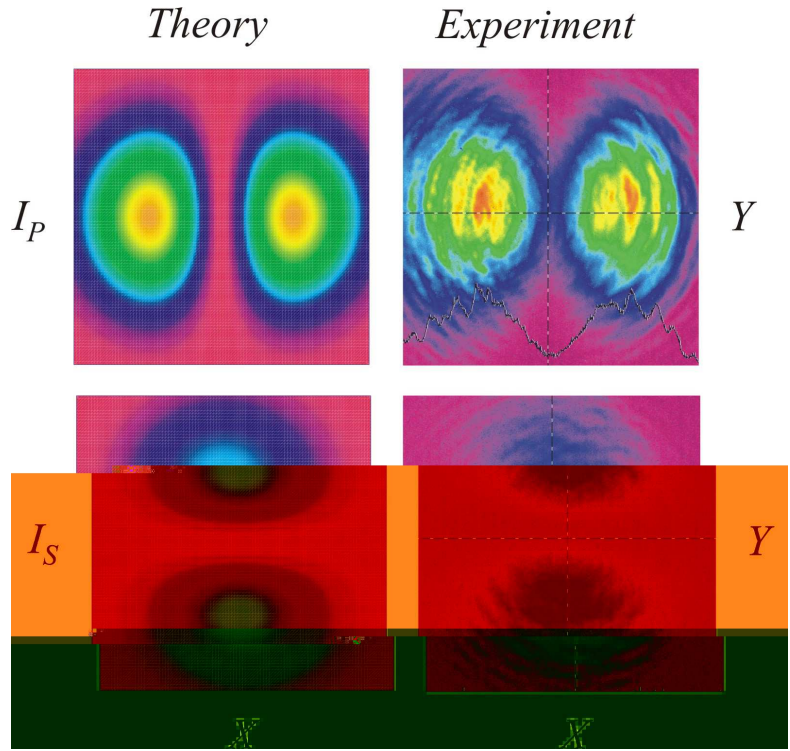}}
  \caption{(Color online) Calculated and measured intensity transverse spatial profiles of the $P$- and $S$-polarized modes of the reflected beam. The beam waist of the incident beam was $w_0 = 34 \, \mu \text{m}$.}
  \end{figure}
  \end{flushright}
%
%
A qualitative comparison between calculated and measured intensity distributions is shown in Fig. 2. The \emph{measured} ratio $\rho_\text{exp}$ of the $S$-polarization component power  to the $P$-polarization one at Brewster incidence, was $\rho_\text{exp} = 0.20 \pm 0.05$ in excellent agreement with the theoretical prediction of Eq. (\ref{OL90}) giving, for BK7 glass, $\rho_\text{th} =  {4 n^4}/{(1 + n^2)^4} \cong 0.18$.
\newpage
In conclusion, we found that when a $TM$-polarized fundamental Gaussian beam is reflected at Brewster incidence it generates a two-mode beam with both a dominant and a cross-polarized component. The intensity of the latter scales quadratically with the angular divergence of the incident beam and can be, therefore, orders of magnitude bigger than the intrinsic cross-polarized intensity of the incident beam that scales with the fourth power of $\theta_0$.

This project is supported by FOM.
\section*{Appendix A}
As a result of the power series expansion truncated at second order terms in $\theta_0$, we found the following expressions for the electric and magnetic fields for the reflected beam, evaluated in the beam frame $K_r$:
\begin{align}
\mathbf{E}^R(\mathbf{r}) = & \, \frac{\exp \left( - i \frac{X^2 + Y^2}{i - Z} \right)}{i - Z}\left( \hat{\mathbf{x}}_r E_{x_r} +  \hat{\mathbf{y}}_r E_{y_r} + \hat{\mathbf{z}}_r E_{z_r} \right), \label{Electric} \\
\mathbf{B}^R(\mathbf{r}) = & \, \frac{\exp \left( - i \frac{X^2 + Y^2}{i - Z} \right)}{i - Z}\left( \hat{\mathbf{x}}_r B_{x_r} +  \hat{\mathbf{y}}_r B_{y_r} + \hat{\mathbf{z}}_r B_{z_r} \right), \label{Magnetic}
\end{align}
where $X = x_r/w_0, \, Y = y_r/w_0, \, Z = (z_r + d)/L$, and
\begin{align}
E_{x_r} = & \, r_P+\theta_0 X\frac{r_P'  }{i-Z }+\frac{\theta_0^2}{2} \Biggl\{i \frac{ (r_P''-2 r_P)}{2 (i-Z )}+i \frac{r_P' \cot \theta }{2( i- Z )}-\frac{i (r_P+r_S) \cot^2 \theta }{ i-Z } \nonumber \\
& \, + X^2 \frac{(r_P''-2 r_P)}{ (i-Z)^2}+Y ^2 \left[\frac{r_P' \cot \theta }{ (i-Z )^2}-\frac{2(r_P+r_S) \cot^2 \theta }{(i-Z)^2}\right]\Biggr\}, \\
E_{y_r} =  & \,  \theta_0 Y  \frac{(r_P+r_S)\cot \theta }{i-Z }+  \theta_0^2 X Y \left[\frac{(r_P'+r_S') \cot \theta }{(i-Z )^2}+\frac{r_S-(r_P+r_S) \csc^2 \theta }{(i-Z )^2}\right],   \\
E_{z_r} =  & \,  \theta_0 \frac{r_PX }{i-Z }+\theta_0^2 \left[i\frac{ r_P'}{2 (i- Z) }+i\frac{ (r_P+r_S) \cot \theta }{2 (i-Z )}+X^2\frac{r_P' }{(i-Z )^2}+ Y^2\frac{(r_P+r_S)  \cot \theta }{(i-Z)^2}\right],
\end{align}
and
\begin{align}
B_{x_r} = &  \, -\theta_0 Y  \frac{( {r_P}+ {r_S}) \cot\theta}{i-Z }+  \theta_0^2  X Y \left[-\frac{( r_P'+ r_S')  \cot\theta}{(i-Z )^2}+\frac{( {r_P}+ {r_S}) \cot^2\theta}{(i-Z )^2}\right], \\
B_{y_r} = &  \,   {r_P} +  \theta_0 X \frac{ r_P'}{i-Z }+ \frac{\theta_0^2}{2} \Biggl\{ i \frac{  r_P'  \cot\theta}{2( i- Z )}+i\frac{ r_P''+2  {r_S}-2 ( {r_P}+ {r_S})  \csc^2\theta}{2 (i-Z )}  \nonumber \\
&  \, + X ^2\frac{( r_P''- {r_P})}{ (i-Z )^2} +Y ^2 \left[\frac{ r_P'  \cot\theta}{ (i-Z )^2}+\frac{ {r_P}+2  {r_S}-2 ( {r_P}+ {r_S}) \csc^2\theta}{ (i-Z )^2}\right] \Biggr\},\\
B_{z_r} = &  \,  \theta_0   Y  \frac{ {r_P}}{i-Z }+   \theta_0^2 X Y \left[\frac{ r_P'}{(i-Z )^2}-\frac{( {r_P}+ {r_S})  \cot\theta}{(i-Z )^2}\right].
\end{align}
The expressions above have been calculated for a $P$-polarized incident beam, and  $r_P$ and $r_S$ are the Fresnel reflection amplitude for $P$ and $S$ waves, respectively, evaluated at the central angle of incidence $\theta = \arccos(\mathbf{k}_0 \cdot \hat{\mathbf{z}}/k_0)$. Moreover, we have used the notation $r_A' = \partial r_A/\partial \theta$ and $r_A'' = \partial^2 r_A/\partial \theta^2$, where $A \in \{ P,S \}$. It is worth noting that the cross-polarization terms at first order in $\theta_0$ present in the expressions of $E_{y_r}$ and $B_{x_r}$, have the same functional dependence $\sim (r_P + r_S)$ given in Eqs. (16) and (19) of Ref. \cite{NaPa}. Therefore, since for the geometrically reflected beam (which is the specular image of the incident one) we have $r_P = 1, \, r_S =-1$, it is clear that first order cross-polarization terms disappear.
\section*{Appendix B}
In Eqs. (\ref{OL50}-\ref{OL60})  $u, \, v, \,  p , \, q$ and $ s $ are some  functions of $Z, \theta, \,r_P, r_S$ and their derivatives. Their explicit form is given below:
\begin{align}
u & \,  = - 2 r_P r_P' \frac{Z}{1 + Z^2} \label{u}, \\
v & \,  =  \frac{r_P \left[ r_P'' + 2 r_S + r_P' \cot \theta - 2\left( r_p +r_S \right) \csc^2 \theta  \right]}{2(1 + Z^2)} \label{v}, \\
p & \,  = \frac{ -r_P (1 - Z^2) \left( 2 r_P'' - 3 r_P\right) + 2 {r_P'}^2 (1 + Z^2)  }{4(1 + Z^2)^2} \label{p}, \\
q & \,  = \frac{ r_P (1 - Z^2) \csc^2 \theta  \left[ 5 r_P + 4 r_S + \left( 3 r_P + 4 r_S\right) \cos 2 \theta - 2 r_P' \sin 2 \theta  \right]}{2(1 + Z^2)^2} \label{q}, \\
s & \,  =  \frac{\left( r_P +r_S \right)^2\cot^2 \theta}{1 + Z^2} \label{s}.
\end{align}
These functions look quite complicated, however $u$, $v$ and $q$ are proportional to $r_P$ so that they disappear when evaluated at Brewster angle of incidence. In passing, we mention here that the linear term $\theta_0 u X$ in Eq. (\ref{OL50}) generates the Goos-H\"{a}nchen shift in the reflected beam.
%
%

\begin{thebibliography}{10}
\bibitem{NoteGH}
In passing, we mention here that the linear term $\theta_0 u X$ in Eq.
  (\ref{OL50}) generates the Goos-H\"{a}nchen shift in the reflected beam,
  where $u = - \frac{2 Z r_P}{1 + Z^2}\frac{\partial r_P}{\partial \theta}$ for
  any angle of incidence.
\end{thebibliography}
%

\end{document}